\documentclass[twocolumn,9pt]{article}

\usepackage{balance}
\usepackage{etoolbox}
\usepackage[export]{adjustbox}
\usepackage[utf8]{inputenc}
\usepackage{flushend}
\usepackage[section]{placeins}
\usepackage[bf, footnotesize]{caption}
\usepackage{graphics}
\usepackage{CJKutf8}

\usepackage{amssymb}
\usepackage{mathtools, cuted}
\usepackage{widetext}

\usepackage{lineno}

\usepackage{multicol}
\usepackage{booktabs}
\usepackage{amsmath}
\usepackage{widetext}
\usepackage{flushend}
\usepackage[explicit]{titlesec}
\titleformat{\section}
  {\normalfont}{\thesection}{1em}{\MakeUppercase{#1}}
  \titleformat{\subsection}
  {\it}{\thesection}{1em}{{#1}}
  
\begin{document}
\small
\begin{strip}

\title{The Turing machine of a harmonic oscillator: from the code to the dynamic system}
\date{}


\author{Francesco Sisini, Valentina Sisini\\
\begin{small}
  Tekamed srl
\end{small}\\\\
\begin{tiny}
Published on (work in progress)
\end{tiny}
}



\begin{center}
\line(1,0){450}
\end{center}
\maketitle

\pagebreak



\section*{Abstract}
In this work we consider a dynamic system consisting of a damped harmonic oscillator and we formalize a Turing Machine whose definition in terms of states, alphabet and transition rules, can be considered equivalent to that of the oscillator.
We prove that the Turing Machine of a FOR loop corresponds to that of the oscillator and we ask ourselves if it is possible to obtain the dynamic system of the harmonic oscillator as a physical realization of the FOR loop. We discuss the relationship between the results found and the \textit{science of Can and Can't}.\\
We discuss the possibility of an evolution of computer science also towards non-computerized specialized machines whose operating principle is designed as an automatic process starting from a source code instead of as a work of human ingenuity.\\The approach to the implementation of algorithms in dynamic systems instead of universal computers can be particularly interesting for the field of both diagnostic and implantable medical devices.

\end{strip}

\section*{Introduction}

Alan Turing formally introduced the concept of computability of a function and proved that a universal automaton (Universal Turing Machine) could be built capable of computing any computable function \cite{Turing}. Although he was not the first to have such an idea \cite{Sinderen}, his work effectively opened the era of computing and computers. \\
The Turing automaton (or Turing Machine, TM) consists of a tape of infinite length on which a head writes or erases symbols belonging to an alphabet. The head can also read symbols and an internal state is defined for it. Based on the state and the symbol read, a series of transition rules specify the next action to be taken by the head, ie what to write, which direction to run the tape and which state to take. The evolution over time of the symbols on the tape depends only on the symbols on the tape at the initial moment of the computation and the transition rules.
A Universal Turing Machine (UTM) is a TM Turing that can compute any computable function. A TM that computes a precise function can also be referred to with the term algorithm.

Over time, technological evolution has led to the creation of several computational automata very different from the UTM and which exploit mechanisms different from the original idea of Alan Turing but all these have always proved to be equipotent to the UTM in the sense that they can compute exactly the same functions \cite{Neumann}. This may suggest that it is impossible to build a computational automaton that is not equivalent to a UTM. This deduction is not entirely correct, although it remains true that to date no built or designed automaton can compute functions that are not computable by a UTM. \\
   The UTM is based on an implicitly accepted assumption that is not completely true in the quantum world, i.e. that there is a one-to-one relationship between the symbol written on the tape in a given position and the symbol read on the tape in the same position.
\\

As we see below, in the quantum world this idea needs to be revised. \\
In the macrosocopic world, where classical physics is king, we can accept that a symbol of the UTM is anything, for example a letter written on a paper tape or the presence of a hole in a card.
Quantum mechanics, on the other hand, describes a well-defined set of entities such as particles and fields, and only these can be used to create the calculating automaton. It goes without saying that the quantum representation of an ink trace on a paper ribbon is theoretically possible, but it is also absolutely useless as the enormous amount of particles present in the system makes it practically impossible to control its entanglement with the surrounding environment, that is, it is impossible to exploit the characteristics of coherence that make quantum systems quantum [Hornberger]. To exploit the peculiarities of quantum mechanics, the sequence of symbols on the TM tape must be represented by a realistic and above all coherent quantum system as a finite set of individually addressable ions could be. \\

In such a world, UTM symbols could be implemented as the quantum state of the ions. In fact, today it is common to understand quantum computation based on qubits, that is, two-level physical systems, such as the spin of a fermion, which are the quantum analogue of bits.

For the principle of superposition of states, it is possible to think that the qubit written on the quantum tape is a superposition of two states, which does not have a classical analogue, in which the bit is 0 or 1, but not both values simultaneously.

The TM selection rules are based on reading the symbol under the head. An implicitly accepted hypothesis is that the reading result is uniquely determined by the symbol present on the tape, for example if the symbol 1 is present on the tape the reading result is 1. In the realization of a TM based on quantum technologies, this hypothesis cannot be implicitly accepted, in fact the reading of the quantum symbol corresponds to a quantum measurement that leads to the collapse of the system in a basic state, therefore the result of the reading is probabilistic rather than deterministic as in the classical case.

Since, as seen above, the computation of a function is based on the transition rules, which are defined for each symbol read on the tape, the uncertainty of the measurement of the symbol involves the violation of the UTM assumptions. In fact, it is shown that such a machine, i.e. a quantum Turing machine (QTM), is substantially equivalent to a probabilistic Turing machine, i.e. the choice of the transition rule to be applied is probabilistically linked to the symbol that is instead read deterministically from the tape.

In the early 80's, the idea of evolving the UTM into a universal quantum TM (UQTM) stimulated the interest of physicists who considered the possibility of creating a quantum computer \cite{Feynman}.
The goal of creating a quantum computer led to the transformation of Alan Turing's hypothesis of the concept of a universal machine to compute functions to that of David Deutsch of a universal dynamic system to simulate dynamic systems.
In Turing's view, the UTM is an abstract automaton that computes abstract functions:
\begin{center}
\textit{Every ‘function which would naturally be regarded as computable’ can be
computed by the universal Turing machine}
\end{center}
Deutsch's vision starts from the principle that a UTM is a physical system and that any physical system can be simulated by the UTM:
\begin{center}
  \textit{
Every finitely realizable physical system can be perfectly simulated by a
universal model computing machine operating by finite means
}
\end{center}

The vision of Deutsch \cite{Deutsch} helped to define what is meant by QTM and UQTM and also served as a trait d’union between the concept of algorithm seen as an abstract procedure and the hardware needed to implement it.

In our opinion, integrating Deutsch's vision with Turing's it can be deduced that every Turing-computable function can be realized as a physical or dynamic system, both on the classical and on the quantum level. This is obviously not an original discovery, but one that is worth emphasizing because its implications are very interesting.

It should be noted that it was precisely the transition from classical to quantum computation that contributed to this new vision. As we have seen, in the quantum world there are elementary two-level systems that lend themselves to being interpreted as qubits so it is easy to understand the execution of an algorithm as the dynamic evolution of these qubits.

With the qubits computation model it is possible to think of a computation automaton as a set of qubits, whose transformations are determined by unitary operators $U$. From the computational point of view, the operators $U$ represent the transition rules of the Turing Machine and therefore it is simultaneously both the program and the dynamic system itself. In these terms, in quantum computing there is no clear difference between hardware and software.

It is natural that this vision also contaminates classical computation where instead the separation between hardware and software is clearer and computation is more identified with the computation of a mathematical function than with the evolution of a dynamic system. 

A common computer can be seen as a classical type dynamic system. Although modern computers make use of technologies based on quantum phenomena, such as semiconductors, it is true that they could also be built with purely classical methods and technologies such as thermionic valves, and there are even examples of functioning computers made entirely of wood. However, in the perspective in which the computer is seen as a dynamic system, we ask ourselves how the software should be seen. \\
An ordinary computer, that is, built following von Neumann's architecture, is composed of a set of registers and a memory. Each register is composed of a certain number of bits, for example 64, each of which can store the symbol 1 and the symbol 0, two abstract symbols that are used to represent two different magnetization states of the bit or two different levels of electrical voltage. The same is true for memory which can be seen as a single register of a much larger size (giga or tera bits).
 
From a physical point of view, therefore, the computer is a physical system whose dynamic evolution can be expressed in the temporal evolution of the physical states of its registers and its memory. In practice, during the execution of a program there are no moving parts (obviously the fans or the pressure of the keys are excluded, which in this context have no meaning) but there is a variation, for example, of the magnetization state of the bits representing internal registers or memory. In this perspective, a program loaded into memory before being executed represents the initial conditions of the dynamic system, that is, a program is the initial condition of the dynamic computer system that uniquely defines its evolution over time.

Each classical computer is therefore a dynamic system that follows the laws of physics, and each program defines the specific initial conditions of this system.

This series of considerations has stimulated our view which consists in considering computable functions in terms of a dynamic system.
  In practice we think that given a function, or the algorithm that implements it, there should always exist a dynamic system, that is, a physical system composed of interacting parts, whose dynamic evolution over time produces the computation encoded in the algorithm, this is in the classical world as well as in the quantum one. This stimulates the interest in identifying an automatic procedure to obtain a dynamic system (output) starting from an algorithm (input). \\
In practice it is a matter of thinking about how an algorithm can be transformed into a dynamic system, therefore into a physical system whose dynamic evolution corresponds to the computation of the algorithm.
In other words, it is a question of creating a physical \textit{particular} or dedicated machine, rather than the \textit{universal} that implements only the desired algorithm.

 This objective is not purely speculative, in fact the realization of dedicated dynamic systems rather than computerized systems could have some advantages.
 From a practical point of view, for example, dedicated dynamic systems could be built with fewer limitations regarding the temperature conditions, ionizing radiations, pressure, etc., compared to similar systems based on the presence of a processor that is classical or quantum. This approach is also reminiscent of Brooks' robots to which he refers when speaking of intelligence without representation [Brooks].
 These systems could find particular application in the field of medical physics both as support devices or prostheses and as diagnostic tools. \\

 In this preliminary article we analyze a simple physical system to derive a Turing Machine whose computation corresponds to the dynamics of the given system. In this context we analyze the conditions for which a classical physical system can be considered an information medium and the principles for deriving the Turing Machine from the dynamic system.

 In this work we limit ourselves to a purely classical analysis, postponing quantum analysis to a moment in which the main obstacles in its realization will have been overcome.   

 \part*{Background}
 
\section*{Turing Machine} 
The Turing machine is an automatic calculation system formed by a tape composed of cells on each of which there can be written a symbol $s \in \Sigma$ and a head that can read and write symbols on the tape. A set $Q$ of states in which the head can be found is also defined for each TM.
The symbols belong to a non-empty finite set $\Sigma$ called the ribbon alphabet. \\

The head moves left or right on the tape and reads and overwrites the cell it is on according to the transition rules ($R$) specific to each TM. The transition rules are applications $r$ like:
$$
r: Q\times\gamma \rightarrow Q\times\Gamma\times\{+,-,0\}
$$
and given in the form:
\begin{equation}
  \langle qi,si,qf,sf,r\rangle
  \label{eq:istruzione}
\end{equation}
where $qi$ indicates the state the TM is in.
Each instruction given in the form \ref{eq:istruzione} specifies: the initial state of the head, the symbol read, the state in which the head is brought, the symbol to be written instead of the symbol read and the direction of movement: right, left or stationary.
 
 \section*{Dynamic systems in classical and non-relativistic mechanics}
 In classical mechanics a dynamic system is described through the concept of a material point.
A material point has three spatial coordinates that define its position with respect to a system of axes and its mass.
 Each dynamic system can be seen as a set of a certain number N of material points which interact with each other, an image not far from the atomic model of matter, in which matter is represented as a set of atoms linked together by forces.
 The interaction between the material points consists of a force that must be a function of their distance and other properties such as the mass or the electric charge of the material points themselves. Force can also be a function of velocities such as in the case of frictional forces or in the non-relativistic description of the Lorentz force.
 Solid systems can also be described as a set of points bound together, such as two masses at the ends of a handlebar of negligible mass. The analysis of the dynamics of a system through the constraints is considerably simplified compared to the more precise analysis that investigates the nature of the constraints themselves. For example, it is much easier to consider a solid as a series of atoms bound together than to consider the force of attraction that exists between the atoms that make up the same. This form of analysis of dynamic systems in terms of their constraints finds its maximum formal expression in what is known as Lagrangian analysis.

 \section*{Harmonic oscillator}
An example of a dynamic system is the harmonic oscillator, this is a very important example in the physical literature as it allows to approximate many real physical systems and was used by both Planck and Einstein to discuss the emission spectrum of the black body.
In practice, two material points are considered, one with a very large mass compared to the other and subject to a force that depends directly on the distance that separates them. The intensity of the force is given by the product of the distance by a constant of proportionality called the elastic constant.
By moving the material point of lower mass to a certain distance from the second material point and then letting it go, a periodic oscillation motion is established whose period of oscillation is given by the square root of the ratio between the elastic constant and the mass. The material point is always accelerated except when it passes through the central position, the instant in which its acceleration is zero.
A real example of a harmonic oscillator is given by a system of two masses connected by a spring that has a certain rest length. By moving the lighter mass $m$ away and then letting it go, a harmonic motion of the smaller mass is established, while it can be considered that the greater $M$ remains at rest.
In the limit in which the mass $M$ is much greater than $m$, the former can be excluded from the analysis and it can be considered that the mass $m$ is attracted towards the coordinate $x = 0$ in which is positioned $M$. Obviously this is only a mathematical device to simplify the treatment, but the results obtained are in any case the same that would be obtained considering both masses, but the treatment is considerably simplified. \\

The Lagrangian function for this system is obtained after writing the expression for kinetic energy $K=\frac{1}{2}m\dot{x}^2$ and for potentail energy  $U=\frac{1}{2}kx^2$.

\begin{equation}
\mathcal{L}=\frac{1}{2}m\dot{x}^2-\frac{1}{2}kx^2
\end{equation}

The equations of motion are given by the Euler-Lagrange equation:

\begin{equation}
\frac{d}{dt}\frac{\partial\mathcal{L}}{d\dot{x}}-\frac{\partial\mathcal{L}}{dx}=0
\end{equation}

They are expressed in terms of the second derivative of the position with respect to time, therefore as the acceleration:
\begin{equation}
\ddot{x}=-\frac{k}{m}x
\end{equation}

The system therefore consists of a material point whose acceleration and position vary harmonically over time:
\begin{equation}
x(t)=A\cos\left(\sqrt{\frac{k}{m}}t\right)
\end{equation}
where the value of $A$
\begin{equation}
  A=x(0)
  \label{eq:ci}
\end{equation}
represents the initial conditions of motion.

\begin{figure}
  \includegraphics[width=0.5\textwidth]{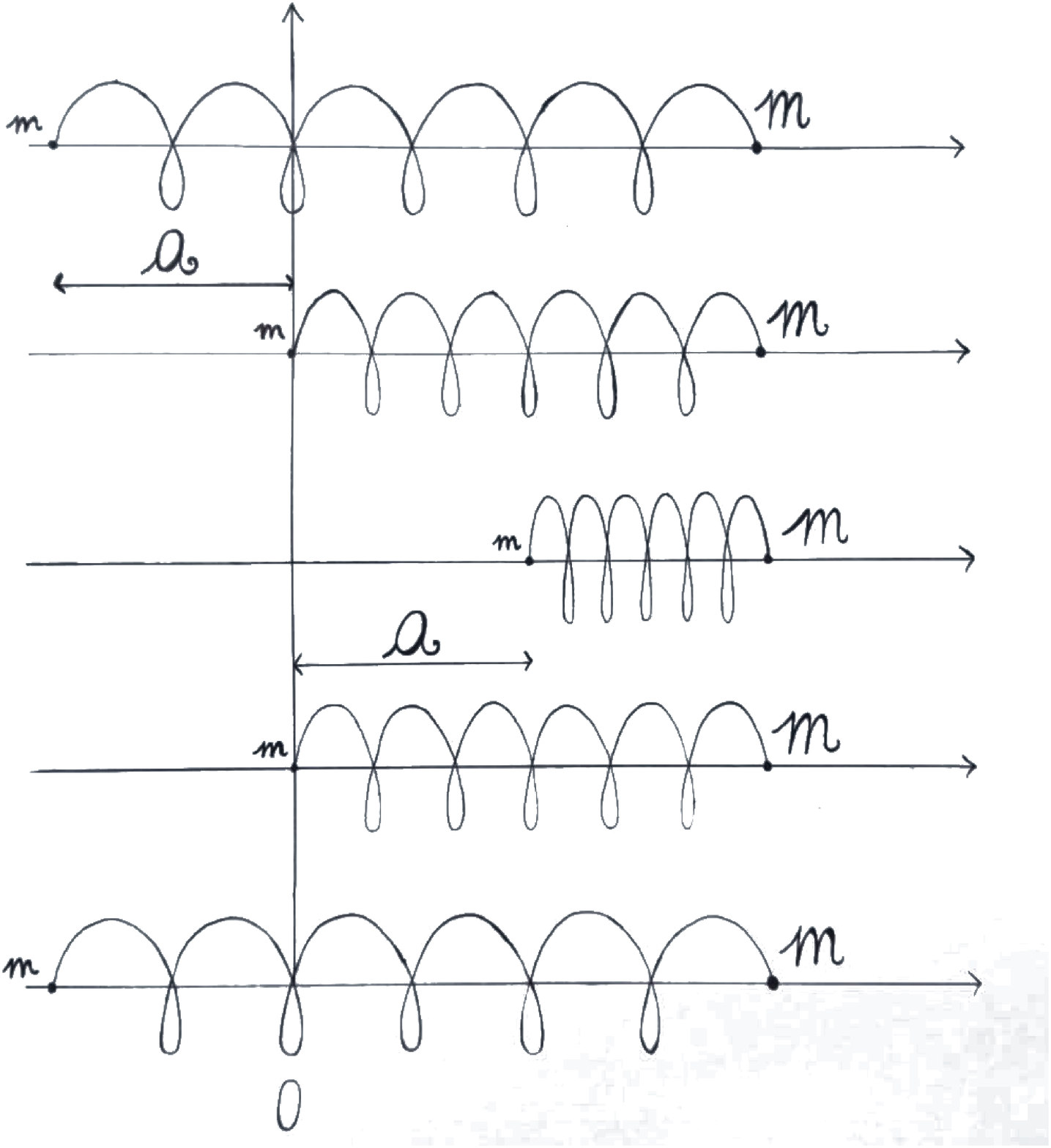}
  \caption{
The figure represents a complete period of the oscillation of a harmonic oscillator. The mass $M$ is assumed to be much greater than the mass $m$.
  }
  \label{fig:AT}
 \end{figure}

 \subsection*{Damping friction}
If it is assumed that the spring offers a certain resistance to motion, it is realized that the oscillations will not be able to persist indefinitely over time but will gradually fade to a stop. The resistance of the spring is inserted in the description of the system by a term of the Lagrangian that depends on the velocity, the equations of motion become:
 
\begin{equation}
\ddot{x}=-\frac{k}{m}x-\mu\dot{x}
\end{equation}

The new equations of motion are characterized by a multiplicative term that decreases with time:
\begin{equation}
x(t)=e^{-\frac{\mu}{2m}t}A\cos\left(\frac{1}{2}\sqrt{4\frac{k}{m}-\frac{\mu^2}{m^2}}t\right)
\end{equation}

Therefore, with each oscillation, the amplitude of oscillation decreases until it is reduced to zero, as is common experience when observing a dynamometer reaching equilibrium or a real pendulum, whose dynamics for small oscillations are the same as the harmonic oscillator. \\
The oscillation period $T$ is given by $T=\frac{2\pi}{\frac{1}{2}\sqrt{4\frac{k}{m}-\frac{\mu^2}{m^2}}}$.
Corresponding to the instants $t_i$ given by:
\begin{equation}
t_i=\frac{i2\pi}{\frac{1}{2}\sqrt{4\frac{k}{m}-\frac{\mu^2}{m^2}}}
\end{equation}
the oscillator completes a complete cycle and the mass $m$ is at the maximum distance $A_i$ which corresponds to the amplitude of oscillation in case of undamped oscillations. For the initial conditions (\ref {eq:ci}) we have that $A_0 = A$ and $A_i$ is given by:
\begin{equation}
  A_i=A_0\left(e^{-\frac{\pi\frac{\mu}{m}}{\frac{1}{2}\sqrt{4\frac{k}{m}-\frac{\mu^2}{m^2}}}}\right)^i
\end{equation}

We define gamma as
$\gamma=A_i/A_0$, that is the ratio between the amplitude at which the system is oscillating at the i-th oscillation and the initial oscillation amplitude $A_0$.
We decide to consider the system stopped when $\gamma<\gamma_0$ (for example $\gamma_0=0.1$)

First of all we see that once the elastic constant, the mass and the coefficient of friction of the system have been assigned, the oscillator will perform a finite and precise number of oscillations $M$ before $\gamma<\gamma_0$:
\begin{equation}
  M=f(\mu)=\sqrt{
    \frac{
      \ln^2(\gamma)\left(4\frac{k}{m}-\frac{\mu^2}{m^2}\right)
    }
         {
           4\pi\frac{\mu^2}{m^2}
           }
  }
  \label{eq:M}
\end{equation}
As the \label{eq:M} equation shows, once the $m$ and $k$ parameters of the system are set, the number of oscillations is only a function of $\mu$.
At the instants $t_i$ defined above, the mass will be at the maximum distance from the center of the oscillation. In practice, the dynamics of the oscillator is reduced to M oscillations of amplitude ($ A_0, ..., A_M $) which all occur with the same duration or period of oscillation, which in fact remains constant.

\subsection*{Free and damped system oscillations}
To give an example of the dynamics described above, consider that the mass $m$ of the system is 1 $kg$, the elastic constant is 100 $N/m$ and the friction coefficient is 0.73 $N/(ms)$.

By setting the initial conditions $x(0)=-5$ the one-dimensional motion of the mass $m$ is represented in the diagram below.
As can be seen, the mass $m$ performs 10 oscillations of amplitude $A_i$ where $i:0.10$. The values of $A_i$ are decreasing and such that $A_i\ gamma_0 > A_0$ for $i <10$ while at the eleventh oscillation, the amplitude $A_10$ is less than one tenth of the initial $A_0$ oscillation.

We have described this example because we will use it later.
\begin{figure}
  \includegraphics[width=0.5\textwidth]{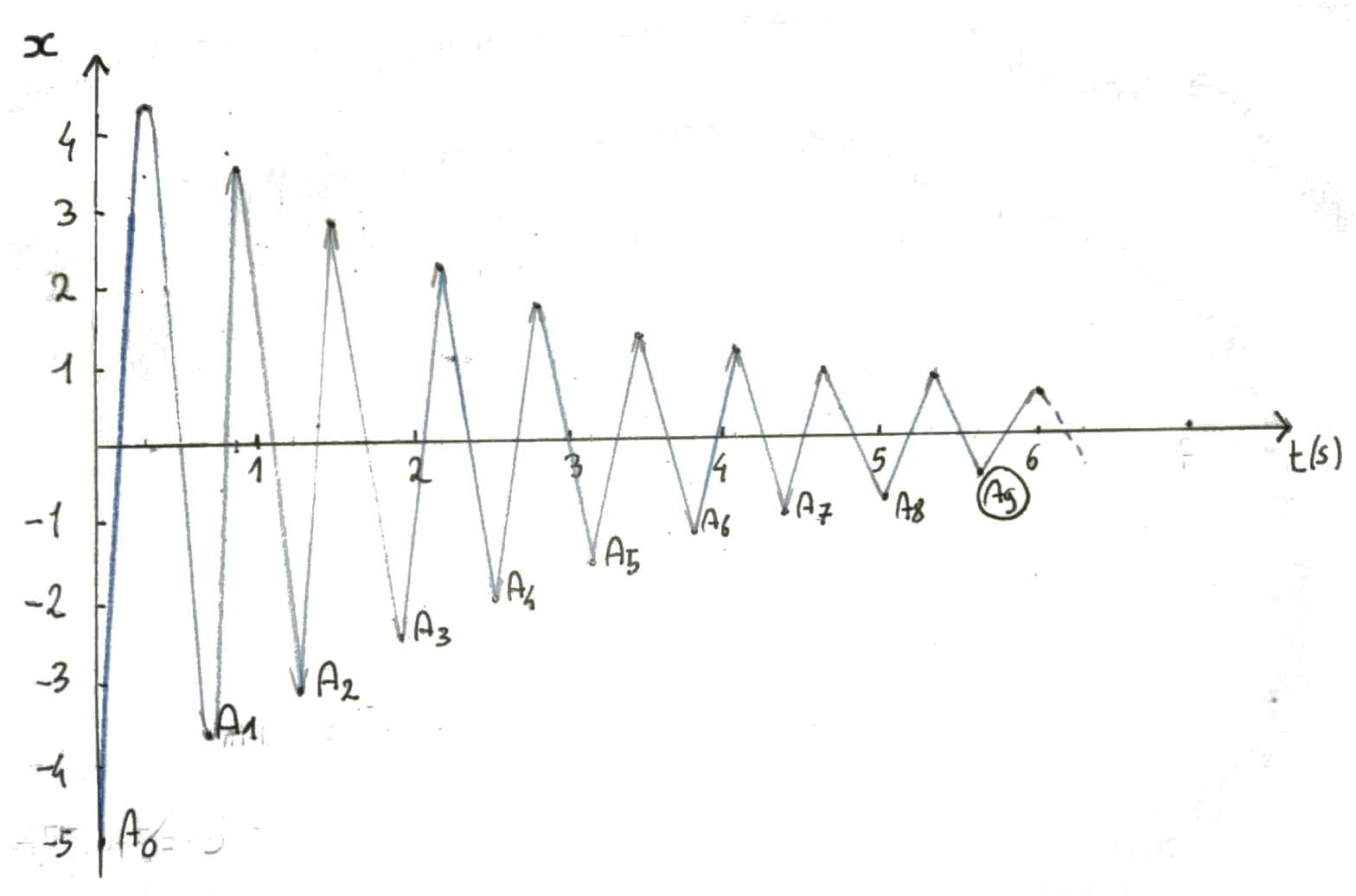}
  \caption{
    The figure represents the dynamic evolution of a damped harmonic oscillator that performs ten oscillations before its oscillation amplitude is reduced to one tenth of the initial value.
  }
  \label{fig:armonico}
 \end{figure}

\part*{Turing Machine of a harmonic oscillator}
\subsection*{Representation of information in classical mechanics}\label{sec:informazione}
Classical dynamical systems are made up of interacting material points whose equations of motion can be deduced once the Lagrangian function of the system has been written.
Based on this we must establish how a message, or rather information in the sense defined by Shannon [Shannon] information theory, can be represented in a dynamic system. \\
Below we will see some examples of systems that do not allow the information to be represented correctly and others that do allow it.

\subsubsection*{Negative examples of information media}
The position of a single material point is not suitable for representing information as one of the fundamental symmetries of classical mechanics is the invariance of the dynamics for spatial translations, to which the conservation of the momentum is associated.
This symmetry derives directly from the idea of homogeneity of space and to date it has never been violated.
In practice it tells us that if a material point is placed in a point in space, if it is isolated from the rest of the universe, its dynamic evolution will be the same and will be indistinguishable from what it would have been if it had been placed in a another point. Each point in the universe is equivalent, so it is not possible to use the position of a single material point as information.
To understand this point, which is very central to the discussion, imagine a fantasy scenario in which the universe has a center. In this context, once a unit of measurement has been set, it would be possible to establish absolute coordinates of each material point with respect to this center of the universe and then use the position of a material point as information. But in classical mechanics, the universe has no center.
Similarly, even the velocity of a single material point cannot be used as a means of information as it depends on the reference system from which the material point is observed.

 \subsubsection*{Positive examples of information media}
One possibility is to use two material points to represent the information that could be encoded in their scalar distance. The distance between two points is in fact invariant under any translation or rotation of the observation point, or rather of the reference system, so the distance between two points could be used as a means of information.
Similarly, the relative velocity between two material points could also be used as a means of information.

In practice we see that a single material point is not a means of information, while a structure, even if minimal, of material points is a means of information.

Another possibility is to encode the information in acceleration. In fact, in classical dynamics, a body that is accelerated in an inertial reference system will also be accelerated to any other inertial observer, therefore also the dynamic state of acceleration can be used to encode information.
Obviously, it should be noted that a material point that is free, therefore not interacting with other material points, cannot present an accelerated motion, so this example also involves the presence of at least two material points that interact by means of a force, or the presence of an external field.

 \subsection*{Encapsulation of information in oscillations}
Compared to the considerations made in the previous paragraph regarding systems suitable to be considered as means of information, we see that the harmonic oscillator can be used in this sense. It is in fact made up of two material points, or a material point and a center of force. The dynamics of the oscillator can be divided into two classes: the oscillation state determined by the condition $A_i/A_0>\gamma_0$ and the rest state s $A_i/A_0\le\gamma_0$. These two classes are physically distinguishable and can be used as a means of information. Oscillation damping can be seen as an information process that transforms the $A_i/A_0>\gamma_0$ state into the $A_i/A_0\le\gamma_0$ state.

 \section*{Turing automaton of the damped harmonic oscillator}
 To derive the Turing Machine of the oscillator we must establish the $Q$ set of states, the $\Sigma$ set of the alphabet and the $R$ set of transitions.
 
 \paragraph*{Q set of states}
We assume that there is a one-to-one correspondence between the $Q$ states of the Turing Machine and the possible states of the system. In the case of the harmonic oscillator it is possible to define the state $q0$ of rest and the state $q1$ of oscillation in the first normal way.
The set $Q$ is therefore given by $Q=\{q0,q1\}$;

\paragraph*{Alphabet}
The configuration of the symbols on the tape fixes the variables of the dynamic system, that is, the initial conditions, the boundary conditions and the free parameters. In the case of the harmonic oscillator, only the friction coefficient $\mu$ must be fixed, while it is not necessary to set the amplitude of the oscillation as an initial condition as the dynamics is independent of it. The set of symbols S must therefore be sufficiently expressive to represent the value of the coefficient of friction. in this case the symbol 1 alone is sufficient, with which any natural number can be represented as a sequence of juxtapositions of the symbol itself.
The tape must therefore contain information that uniquely determines the mu coefficient. Our choice is to represent on the ribbon the value of the expression:
….
That fixed the mass m and the elastic constant of the oscillator is an invertible function of $\mu$.

\paragraph*{Transition rules}
The transition rules define what can happen during dynamic evolution. In the case of the harmonic oscillator, each complete oscillation corresponds to a cyclic sequence of transformations of potential energy into kinetic energy and vice versa.
Our choice is that there are two transition rules, one that allows you to pass from a state of oscillation to a new state of oscillation if the symbol 1 is present under the head of the Turing Machine, and the other that stops the oscillations if there is a null symbol under the head: $\langle q1,1,q1,_,-\rangle$, $\langle q1,_,q0,_,0\rangle$

\subsection*{Computing with damping}
To perform a Turing Machine computation, we set the same parameters set previously: $m$ and $k$ and assume $\gamma=0.1$. The value of the friction coefficient calculated by inverting the \ref{eq: M} results in $(\mu=0.73)$ and determines the value of $f(\mu)=10$ which is then represented on the tape as a string of 10 consecutive 1 digits (see figure).
At the beginning of the computation, the head is located above the symbol 1, so the $\langle q1,1, q1, _, - \rangle$ rule is applied, effectively consuming a possibility of oscillation. The computation continues for 10 iterations until the head finds the blank symbol under it, goes to the state q0 according to the $\langle q1, _, q0, _, 0\rangle$ and ends the computation since no transitions are foreseen for the symbol blank.

\subsection*{Computing without damping}
To perform a Turing Machine computation, we set $\mu=0$ which determines the value of $f(\mu) = \infty$ which is then represented on the tape as an infinite string of consecutive 1 digits.
At the beginning of the computation, the head is above the symbol 1, hence the rule $\langle q1,1,q1,_,-\rangle$ is applied, effectively consuming a possibility of oscillation. Computation continues indefinitely.

\subsection*{Computation analysis}
In the damped case, the Turing Machine makes 10 transitions in the q1 state, which correspond to 10 oscillations. At each oscillation a digit 1 is canceled, therefore a possibility of oscillation is consumed, that is a possible conversion of potential energy into kinetic energy. The idea of consumption of the possibility of oscillation is consistent with the phenomenon of energy dissipation due to the friction present in the system. When all 1s have been removed from the tape the computation stops.
In the ideal case without damping, there are infinite 1 symbols on the tape so the computation never stops. From a physical point of view it means that it is possible to convert potential energy into kinetic energy and vice versa infinite times.

\begin{verbatim}

_|1|1|1|1|1|1|1|1|1|1|q1|1|_ <q1,1,q1,_,->

_|1|1|1|1|1|1|1|1|1|q1|1|_|_ <q1,1,q1,_,->

_|1|1|1|1|1|1|1|1|q1|1|_|_|_ <q1,1,q1,_,->

_|1|1|1|1|1|1|1|q1|1|_|_|_|_ <q1,1,q1,_,->

_|1|1|1|1|1|1|q1|1|_|_|_|_|_ <q1,1,q1,_,->

_|1|1|1|1|1|q1|1|_|_|_|_|_|_ <q1,1,q1,_,->

_|1|1|1|1|q1|1|_|_|_|_|_|_|_ <q1,1,q1,_,->

_|1|1|1|q1|1|_|_|_|_|_|_|_|_ <q1,1,q1,_,->

_|1|1|q1|1|_|_|_|_|_|_|_|_|_ <q1,1,q1,_,->

_|1|q1|1|_|_|_|_|_|_|_|_|_|_ <q1,1,q1,_,->

_|q1|1|_|_|_|_|_|_|_|_|_|_|_ <q1,1,q1,_,->

_|q0|_|_|_|_|_|_|_|_|_|_|_|_ <q1,_,q0,_,0>

_|q0|_|_|_|_|_|_|_|_|_|_|_|_
\end{verbatim}

\section*{Turing automaton of the FOR loop}
Iteration algorithm. \\
In the field of computer science and the theory of formal languages, we consider the problem of having to repeat a certain operation for a certain number M of times, that is, to re-iterate an operation. A solution to this problem is obtained by using a counter whose value, starting from M, is scaled at each iteration. At the end of the iteration, the counter value is compared with the number 1. If the counter value is greater than 1 then the operation is repeated, otherwise it ends.

The algorithm can be written in a language similar to BASIC as follows:
\begin{verbatim}
10 t := M
20 t := t-1
30 //Fai quello che ti serve
40 if t>1 goto 20
50 end
\end{verbatim}
Where the variable t represents the counter.
Turing automaton of the iteration algorithm
This algorithm can also be represented by a TM. You can choose to represent the number of iterations M to be performed by writing the succession of numbers from 1 to M on the tape as shown in the diagram above.
The alphabet S of the automaton is therefore made up of numbers ranging from 1 to M. For this automaton it is sufficient to define two states of the head that we call q1 and q0 and two transition rules. At the beginning of the iterations, the head is in state q1 above the last written symbol. The first rule states that if the head is in state q1 and the symbol under it is 1 (otherwise the cell is empty), then the head must go to state q1 (it remains unchanged) delete the contents of the cell, and move down one cell to the right.
The second transition rule states that if the state is q1 and the symbol under the head is empty, then the head must go to state q0 and delete the contents of the cell.
As you can see, there is no rule that establishes what to do if the head is in state q0, so the computation stops, exactly as it happens in line 50 of the code seen above.
From the definition of this automaton it is easy to see that set for example M = 10, the computation of the Turing Machine is exactly the same as that reported in the diagram ...
In this sense we can say that the damped harmonic oscillator has the same Turing Machine of the FOR cycle and ask ourselves if we can think of the harmonic oscillator as a concrete implementation of said algorithm.

\subsection*{Compiling the code in the dynamic system}
In the previous paragraphs we have seen that the turing automaton related to a specific algorithm and the turing automaton related to a harmonic oscillator are equivalent, in the sense that they are defined by the same alphabet for the tape, by the same set of states for the head and from the same set of transition rules.
In practice, the harmonic oscillator dynamic system in its dynamic evolution transforms information in the same way that the algorithm coded in Basic does.

Programs that are run from a computer are sequences of 1s and 0s.
it is known that programmers do not code their algorithms directly in sequences of 1s and 0s but using formal languages closer to natural language.
The code of an algorithm written in a formal language, for example in the BASIC language, to be executed must first be compiled in machine language, then in sequences of 1s and 0s.
The compilation must ensure that the execution of the code written in machine language produces the same result as the code written in BASIC starting from the same input.
One way to comply with this guarantee is that the machine code and the BASIC code are both equivalent to the same Turing automaton, therefore that both codes have the same alphabet, the same set of states and the same transition rules.

In this sense we can think of the harmonic oscillator as the result of a compilation of the FOR algorithm which results in the dynamic oscillator system.

Let's consider a hypothetical automatic process that accepts as input the BASIC code of the iteration algorithm considered above and that outputs the lagrangian of the harmonic oscillator. For all practical purposes this process has compiled a code expressed in formal language in a real dynamic system.
In this sense we speak of compilation.

Let us now consider the execution of the algorithm that occurs by setting M = 10.
From Turing's point of view, computation is a sequence of transitions where each transition is a passage from one configuration to another mediated by a transition rule. Each computation is therefore the result of the application of the transition rules.

From a dynamic point of view the transitions seen in the paragraph ... specify what can happen and what cannot happen, for example if there are 1 symbols on the tape and the system is oscillating, then the transition rules prevent it from stopping before we have consumed all 1 symbols. In this sense, the transition rules specify what can happen and what cannot happen. This point seems to us very close to the science of what can be and what cannot be of DD. and CM [Marletto].

An interesting development of this research would be to identify a strategy to automatically switch from the algorithm to the dynamic system. A possible lineup is the following:

\begin{itemize}
\item The TM of the algorithm is produced
\item Q states are identified with a set of dynamic variables
\item We consider all the possible laws of motion given by all the possible applications of the transition rules for the system of dynamic variables  
\item The system of differential equations is identified which has the laws of motion found as solutions
\item The system of differential equations is identified which has the laws of motion found as solutions
 \end{itemize}

\section*{Discussion}
Following the evolution of quantum computing, a question that comes naturally is whether computer science has become a thing for physicists or whether physics is a matter for computer scientists. The developments of the quantum computer have shuffled the cards a lot and in this evolving scenario, Chiara Marletto underlined the natural role played by the quantum nature of subatomic physics in the definition of information and formulated a vision around it known as constructor theory. \\
On the one hand it has been highlighted how the classical theory and the quantum information theory differ from each other, on the other hand they have in common the correspondence between algorithm and dynamic system. \\

\subsection*{Consideration from the perspective of the science of can and can't}
The Turing Machine derived for the harmonic oscillator has highlighted the relationship between the transition rules (set $R$) defined for the automaton and what can happen at the physical level. As we have seen the rule $\langle q1,1, q1, _, - \rangle$ says that if the system has made a complete swing, and there is still energy available, then the system can make a new complete swing. In the case of the free oscillator, without damping, this frames the system as a work-media and therefore an information-media. The analysis of this system seems to us to be framed in the context of the \textit{science of can and can't} and also the result that the oscillator can be seen as an information-media, although obtained at the end of a logical procedure different from the one followed by Marletto and Deutsch seems to us to be consistent. \\
We find another element of coherence that emerges as we pass from the Galilean analysis to the relativistic one (Lorentz invariant). Let's take a step back and review the information coding system we have agreed on for the oscillator system. It was established that the two information-bearing states were: oscillation state $A_i/A_0>\gamma_0$ and non-oscillation states $A_i/A_0<\gamma_0$. If we consider the presence of static friction, we know from experience that under a certain $\gamma$ the oscillation will truly end, therefore in a less formal way, we can establish the oscillation state as $A_i>0$ and the non-oscillation state like $A_i = 0$. With this new definition, we are tempted to associate a non-binary information to the value of $A$ but rather analogic, in fact it could be associated with the oscillator the information $A_i$ itself, in practice the information stored in the oscillator would be the same amplitude of oscillation.
This choice seems correct in a non-Lorentz invariant system, but obviously it is not if such invariance is required, because it goes without saying that the content of the information itself would not be an invariant. \\
It must be said that the Lorentz request for invariance would also have consequences on the very definition of the dynamics of the system, which could not be defined only in terms of the Galilean distance between the two masses. \\
Of particular interest is the case of the damped oscillator. From the computational point of view it can be seen as a work-media with side-effects. Therefore, analyzing it from the point of view of the Turing Spot, the damped oscillator is an information-media, but during its computation the energy of its universe decreases progressively, until it reaches a point where it is no longer usable. \\
The considerations drawn here do not yet have a practical value, but we seemed interested in comparing our approach to the problem with the more solid one outlined by Constructor Theory and we believe that a comparison in these terms can be constructive. \\\\
An interesting case to analyze is that of a set of $N$ oscillators of which only one is in an excited state, while the others are in a rest state. If the oscillators are placed in contact, the excited state can be propagated or \textit{copied} and more than one of them can pass from the rest state to the excited state. Of course, by the energy conservation theorem, the oscillator amplitude of an oscillator decreases as that of the oscillator it comes into contact with increases, so during the process, the oscillator amplitude of each new oscillator will decrease. as long as the information \textit{copy} process stops. It follows that information can be copied from one system to another but this has the side effect of dampening the oscillation, so we can think that the process of copying information requires energy. If, on the other hand, we consider a system composed always of N oscillators, but we limit their interaction to the cases in which once they come into contact, one oscillator stops while the other is excited, then we see that in principle the process can continue to Infinity. These observations are also not new and their germ can already be found in the seminal work of Bennet \cite{bennet}.\\\\
We find our research close to the work of horsman and co authors about how to tell if a given physical system is acting as a computer or not\cite{Horsman}.
\subsection*{Biomedical applications}
 Environmental conditions, such as temperature, humidity, or the level of both ionizing and non-ionizing radioations can affect the operation of electronic devices such as a computer. Because of this, the approach to the implementation of algorithms in dynamic systems instead of universal computers can be particularly interesting for the field of both diagnostic and implantable medical devices.

\section*{Conclusion}
In this paper we have presented the possibility of considering a new type of compilation whose result is not the definition of the code executable by a computer, but the Lagrangian of a particular (specific) dynamic system whose dynamic execution corresponds to the execution of the code to be part of a computer. \\

  the proposed example has introduced the problem of compiling an algorithm in a dynamic system using a very simple example for which it is possible to identify also some automatic production rules.
  the definition of formal rules in general, however, is certainly very complex even if in principle it is to be expected that it will be possible.\\

\pagebreak

\end{document}